\documentclass[12pt]{article}

\begin{document}

\bigskip\ 

\bigskip\ 

\begin{center}
\textbf{ISSUES OF DUALITY IN ABELIAN GAUGE THEORY }

\smallskip\ 

\textbf{AND IN LINEARIZED GRAVITY}

\smallskip\ 

J. A. Nieto$^{\star }$ \footnote{%
nieto@uas.uasnet.mx} and E. A. Le\'{o}n$^{\ast }$ \footnote{%
ealeon@posgrado.cifus.uson.mx}

\smallskip\ 

$^{\star }$\textit{Facultad de Ciencias F\'{\i}sico-Matem\'{a}ticas de la
Universidad Aut\'{o}noma} \textit{de Sinaloa, 80010, Culiac\'{a}n Sinaloa, M%
\'{e}xico.}

$^{\ast }$\textit{Departameto de Investigaci\'{o}n en F\'{\i}sica de la
Universidad de Sonora, Hermosillo Sonora , M\'{e}xico}

\bigskip\ 

\bigskip\ 

\textbf{Abstract}
\end{center}

We start by describing two of the main proposals for duality in Abelian
gauge theories, namely $F$(ield strength)-duality approach and the $S$%
-duality formalism. We then discuss how $F$-duality and $S$-duality can be
applied to the case of linearized gravity. By emphasizing the similarities
and differences between these two type of dualities we explore the
possibility of combining them in just one duality formalism.\ 

\bigskip\ 

\bigskip\ 

\bigskip\ 

\bigskip\ 

\bigskip\ 


Pacs numbers: 04.60.-m, 11.25.Tq, 11.15.-q, 11.30.Ly

May, 2009

\newpage

\noindent \textbf{1.- Introduction}

\smallskip\ 

Duality in linearized gravity [1] has been a topic of considerable interest
[2]-[29]. There are at least two physical reasons for this increasing
interest of the topic. The first possibility arises from the hope of
determining the strong coupling limit for linearized gravity (see Refs. [1]
and [2]) via the analogue of the $S$-duality concept [30] in gauge field
theories. In fact, just as in a dual gauge theory the coupling exchange $%
g^{2}\rightarrow 1/g^{2}$ describes a basic dual symmetry, one may expect a
dual gravitational theory with either one of the exchanges $%
l_{p}^{2}\rightarrow 1/l_{p}^{2}$ [2] or $\Lambda \rightarrow 1/\Lambda $
[1], [26], where $l_{p}$ is the Planck length and $\Lambda $ is the
cosmological constant.

The second motivation comes from the idea of implementing a dual symmetry of
the linearized gravitational field equations at the level of the
corresponding action [5]. Such a dual symmetry is the\ gravitational
analogue of the corresponding electromagnetic dual symmetry provided by the
electric and magnetic field strengths. In this case, the Riemann tensor and
its dual play the role of the electric and magnetic fields strengths
respectively. This dual gravitational approach has its origins in the old
observation [31] that in the case of electromagnetism such a kind dual
symmetry can be implemented at the level of the action if the infinitesimal
transformations are applied canonically to the gauge field rather than to
the corresponding field strength.

From the above comments we observe that while in the $S$-duality approach
[30] the emphasis is put in the coupling exchange, in the case of the
canonical approach the attention is focused on the dual transformation of
the field strength. Both generalized approaches have, however, a common
origin, namely the dual symmetry of the Maxwell equations discovered by
Dirac itself [32-33]. Since linearized gravity can be understood as an
Abelian gauge theory [26] one becomes motivated to see whether there is a
kind of dual theory for gravity in which both coupling and field strength
dual exchanges are equally important. In order to find such a dual
gravitational theory we first need to analyze carefully the differences
between the $F$-duality (field strength duality) and $S$-duality in an
abelian gauge field theory. For this purpose in sections 2 and 4 we briefly
discuss the $F$-duality approach of references [31] and [5], respectively.
In sections 3 and 5, we briefly review the $S$-duality theory for Abelian
gauge fields proposed in Ref. [30] and the $S$-duality theory for linearized
gravity described in Ref. [1], respectively. With this reviews at hand in
sections 6 and 7, we propose a unify duality theory for Abelian gauge field
theory and linearized gravity, respectively. Finally, in section 8 we make
some final remarks.

\bigskip\ 

\noindent \textbf{2. }$F$\textbf{-duality for an Abelian gauge field theory}

\smallskip\ 

In this section, we summarize the main duality ideas of the approach
proposed in Ref. [31]. Consider the field strength $F^{\mu \nu }=-F^{\nu \mu
}$ and its dual%
\begin{equation}
^{\ast }F^{\mu \nu }=\frac{1}{2}\varepsilon ^{\mu \nu \alpha \beta
}F_{\alpha \beta },  \label{1}
\end{equation}%
where $\varepsilon ^{\mu \nu \alpha \beta }$ is the completely antisymmetric
Levi-Civita density in a Minkowski spacetime. The source-free Maxwell
equations are

\begin{equation}
\partial _{\nu }F^{\mu \nu }=0  \label{2}
\end{equation}%
and

\begin{equation}
\partial _{\nu }^{\ast }F^{\mu \nu }=0.  \label{3}
\end{equation}%
It is straightforward to see that these field equations are invariant under
the transformation

\begin{equation}
\delta F^{\mu \nu }=\beta ^{\ast }F^{\mu \nu }  \label{4}
\end{equation}%
and%
\begin{equation}
\delta ^{\ast }F^{\mu \nu }=-\beta F^{\mu \nu },  \label{5}
\end{equation}%
where $\beta $ is an arbitrary constant. Here we used the fact that $^{\ast
\ast }F^{\mu \nu }=-F^{\mu \nu }.$

Since

\begin{equation}
F^{\mu \nu }\delta F_{\mu \nu }=\beta F^{\mu \nu \ast }F_{\mu \nu },
\label{6}
\end{equation}%
the action

\begin{equation}
S_{I}=\frac{1}{2}\int d^{4}xF^{\mu \nu }F_{\mu \nu }  \label{7}
\end{equation}%
is not invariant under (4) unless we write

\begin{equation}
F_{\mu \nu }=\partial _{\mu }A_{\nu }-\partial _{\nu }A_{\mu },  \label{8}
\end{equation}%
which means solving (3). The authors of Ref. [31] pointed out that this
contradictory invariance can be solved if one considers consistent canonical
variations of the potential $\delta A_{\mu }$ instead of variations of the
field strength $\delta F_{\mu \nu }$. With the idea of emphasizing the
invariance of the action (7) at the level of the field strength $F_{\mu \nu
} $ according to (4), we shall refer this approach as $F$-duality formalism.

\bigskip\ 

\noindent \textbf{3. \ }$S$\textbf{-duality for an Abelian gauge field theory%
}

\smallskip\ 

Here, we shall briefly review the $S$-duality formalism for an Abelian gauge
theory (see Ref. [30]). Our starting point is the action

\begin{equation}
S_{II}=\frac{1}{2g^{2}}\int d^{4}x{}F^{\mu \nu }{}F_{\mu \nu }+\frac{\theta 
}{2}\int d^{4}x{}F^{\mu \nu }{}^{\ast }F_{\mu \nu }.  \label{9}
\end{equation}%
Here, it is assumed that $F_{\mu \nu }=\partial _{\mu }A_{\nu }-\partial
_{\nu }A_{\mu }.$ The $\theta $-term is topological and, of course,
classically it can be dropped from (9). This implies that in this case (9)
can be reduced to the action (7). However, if our goal is to quantize the
theory described by (9) it becomes necessary to keep the $\theta $-term.
Observe that in contrast to the formalism of section 2, in this approach
there is an emphasis in the role played by of the constants $g^{2}$ and $%
\theta $.

Now, by introducing the (anti) self-dual field strengths

\begin{equation}
^{\pm }F^{\alpha \beta }=(\frac{1}{2})^{\pm }N_{\tau \lambda }^{\alpha \beta
}F^{\tau \lambda },  \label{10}
\end{equation}%
where

\begin{equation}
^{\pm }N_{\tau \lambda }^{\alpha \beta }=\frac{1}{2}(\delta _{\tau \lambda
}^{\alpha \beta }\mp i\varepsilon _{\,\quad \tau \lambda }^{\alpha \beta }),
\label{11}
\end{equation}%
with $\delta _{\tau \lambda }^{\alpha \beta }=\delta _{\tau }^{\alpha
}\delta _{\lambda }^{\beta }-\delta _{\tau }^{\beta }\delta _{\lambda
}^{\alpha }$ denoting a generalized delta, one can prove that the action (9)
can be written as

\begin{equation}
S_{III}=\frac{1}{2}(\tau ^{+})\int d^{4}x{}^{+}F^{\mu \nu }{}^{+}F_{\mu \nu
}{}+\frac{1}{2}(\tau ^{-})\int d^{4}x{}^{-}F^{\mu \nu }{}^{-}F_{\mu \nu },
\label{12}
\end{equation}%
where $\tau ^{+}$ and $\tau ^{-}$ are two different constant parameters
given by

\begin{equation}
\tau ^{+}=\frac{1}{g^{2}}+i\theta  \label{13}
\end{equation}%
and

\begin{equation}
\tau ^{-}=\frac{1}{g^{2}}-i\theta .  \label{14}
\end{equation}%
The fact that the parameters $\tau ^{+}$ and $\tau ^{-}$ are complex means
that, in addition to the field strength duality transformation,

\begin{equation}
\delta ^{\pm }F^{\alpha \beta }={}\pm i\beta ^{\pm }F^{\alpha \beta },
\label{15}
\end{equation}%
one can in principle implement, for $a,b,c,d\in Z$, the more general duality
transformation

\begin{equation}
\tau ^{\prime }=\frac{a+c\tau }{b+d\tau }.  \label{16}
\end{equation}%
Observe that (16) generalizes the coupling duality transformation 
\begin{equation}
g^{2}\rightarrow \frac{1}{g^{2}}.  \label{17}
\end{equation}%
In fact, it is known that the modular group described by (16) can be
generated by the elements $T:\tau \rightarrow \tau +1$ and $S:\tau
\rightarrow -\frac{1}{\tau }$ (see section 1.4.3 of Ref. [34]). So, if the
vacuum angle $\theta $ vanishes, the $S-$symmetry yields precisely the
transformation (17) .

The next step it is to write a meaningful action which may allow us to
transfer information from the action (9) to its associated dual action.
First, one considers the generalized field strength

\begin{equation}
H^{\mu \nu }=F^{\mu \nu }-G^{\mu \nu },  \label{18}
\end{equation}%
where $G^{\mu \nu }$ is an auxiliary two-form. Secondly, one introduces the
dual field strength $W_{\mu \nu }=\partial _{\mu }V_{\nu }-\partial _{\nu
}V_{\mu }$, where $V_{\mu }$ is a one-form vector gauge field. The
generalized action is then written as [30]

\begin{equation}
\begin{array}{c}
S_{IV}=\frac{1}{2}(\tau ^{+})\int d^{4}x{}^{+}H^{\mu \nu }{}^{+}H_{\mu \nu
}{}+\frac{1}{2}(\tau ^{-})\int d^{4}x{}^{-}H^{\mu \nu }{}^{-}H_{\mu \nu } \\ 
\\ 
+\int d^{4}x{}^{+}W^{\mu \nu }{}^{+}G_{\mu \nu }{}-\int d^{4}x{}^{-}W^{\mu
\nu }{}^{-}G_{\mu \nu }.%
\end{array}
\label{19}
\end{equation}%
This action is invariant under the transformations

\begin{equation}
\begin{array}{c}
\delta A=B, \\ 
\\ 
\delta G=dB,%
\end{array}
\label{20}
\end{equation}%
where $B$ is any one-form. If we eliminate $V$ from (19) one sees that $dG=0$
and therefore we can set $G=0$. Hence from (18) one sees that $H^{\mu \nu
}=F^{\mu \nu }$ and consequently the action (19) is reduced to (12). On the
other hand the gauge invariance (20) allows to set $A=0$ and therefore the
action (19) becomes

\begin{equation}
\begin{array}{c}
S_{IV}=\frac{1}{2}(\tau ^{+})\int d^{4}x{}^{+}G^{\mu \nu }{}^{+}G_{\mu \nu
}{}+\frac{1}{2}(\tau ^{-})\int d^{4}x{}^{-}G^{\mu \nu }{}^{-}G_{\mu \nu } \\ 
\\ 
+\int d^{4}x{}^{+}W^{\mu \nu }{}^{+}G_{\mu \nu }{}-\int d^{4}x{}^{-}W^{\mu
\nu }{}^{-}G_{\mu \nu }.%
\end{array}
\label{21}
\end{equation}%
Finally, after eliminating $^{\pm }G$ one finds that (21) leads to%
\begin{equation}
S_{V}=\frac{1}{2}(-\frac{1}{\tau ^{+}})\int d^{4}x{}^{+}W^{\mu \nu
}{}^{+}W_{\mu \nu }{}+\frac{1}{2}(-\frac{1}{\tau ^{-}})\int
d^{4}x{}^{-}W^{\mu \nu }{}^{-}W_{\mu \nu },  \label{22}
\end{equation}%
which is the dual action. We observe that the coupling constant $\tau $
transforms as $-\frac{1}{\tau }$. Actually, when quantum topological effects
are considered the $\tau $ transformation can be extended to the more
general duality transformation given in (16) (see Ref. [30]).

\bigskip\ 

\noindent \textbf{4.- }$F$\textbf{-duality for linearized gravity}

\smallskip\ 

The Riemann tensor for linearized gravity is given by

\begin{equation}
R_{\mu \nu \alpha \beta }=\frac{1}{2}\left( \partial _{\mu }\partial _{\beta
}h_{\nu \alpha }-\partial _{\mu }\partial _{\alpha }h_{\nu \beta }-\partial
_{\nu }\partial _{\beta }h_{\mu \alpha }+\partial _{\nu }\partial _{\alpha
}h_{\mu \beta }\right) .  \label{23}
\end{equation}%
Here, the object $h_{\mu \nu }=h_{\nu \mu }$ can be understood as a small
deviation from the full metric $g_{\mu \nu }$, namely%
\begin{equation}
g_{\mu \nu }=\eta _{\mu \nu }+h_{\mu \nu },  \label{24}
\end{equation}%
where%
\begin{equation}
(\eta _{\mu \nu })=diag(-1,1,1,1)  \label{25}
\end{equation}%
is the Minkowski flat metric. The vacuum Einstein equations are

\begin{equation}
R_{\nu \beta }=0,  \label{26}
\end{equation}%
where $R_{\nu \beta }=\eta ^{\mu \alpha }R_{\mu \nu \alpha \beta }$ is the
linearized Ricci tensor.

Let us now introduce the dual tensor

\begin{equation}
^{\ast }R_{\mu \nu \alpha \beta }=\frac{1}{2}\varepsilon _{\mu \nu \sigma
\rho }R_{\alpha \beta .}^{\sigma \rho }  \label{27}
\end{equation}%
We observe that due to the Bianchi identity $R_{\mu \nu \alpha \beta
}+R_{\mu \beta \nu \alpha }+R_{\mu \alpha \beta \nu }=0,$ we have that $%
^{\ast }R_{\nu \beta }=\eta ^{\mu \alpha \ast }R_{\mu \nu \alpha \beta }$
satisfies the dual field equation

\begin{equation}
^{\ast }R_{\nu \beta }=0  \label{28}
\end{equation}%
or

\begin{equation}
\frac{1}{2}\varepsilon _{\mu \nu \sigma \rho }\eta ^{\mu \alpha }R_{\alpha
\beta }^{\sigma \rho }=0.  \label{29}
\end{equation}

It is not difficult to see that both field equations (26) and (28) are
invariant under the infinitesimal rotations

\begin{equation}
\delta R_{\mu \nu \alpha \beta }=\beta ^{\ast }R_{\mu \nu \alpha \beta }
\label{30}
\end{equation}%
and

\begin{equation}
\delta ^{\ast }R_{\mu \nu \alpha \beta }=-\beta R_{\mu \nu \alpha \beta },
\label{31}
\end{equation}%
where $\beta $ is again a constant. Comparing the development of section 2
with the present section we observe that these transformations are
completely analogous to the expressions (4) and (5). Thus, it is expected
that the Pauli-Fierz action

\begin{equation}
S_{VI}=4\int d^{4}x(\partial ^{\alpha }h^{\mu \nu }\partial _{\alpha }h_{\mu
\nu }-2\partial _{\mu }h^{\mu \nu }\partial _{\alpha }h_{\nu }^{\alpha
}+2\partial ^{\mu }h\partial ^{\nu }h_{\mu \nu }-\partial ^{\alpha
}h\partial _{\alpha }h),  \label{32}
\end{equation}%
where $h=h_{\alpha }^{\alpha }$, is not invariant under (30) and (31) unless
we describe an infinitesimal canonical transformations in terms of the
potential $\delta h_{\mu \nu }$ instead of the field strengths $R_{\mu \nu
\alpha \beta }$ and $^{\ast }R_{\mu \nu \alpha \beta }$. Actually, the $%
SO(2) $ rotations are achieved by means of two superpotentials; one
associated with $h_{\mu \nu }$ and the other with its canonical conjugate
momenta (see Ref. [5] for details).

\bigskip\ 

\noindent \textbf{5.- }$S$\textbf{-duality for linearized gravity}

\smallskip\ 

Let us start observing that the curvature Riemann tensor $R_{\mu \nu \alpha
\beta }$ for linearized gravity, given in (23), can be written as

\begin{equation}
R_{\mu \nu \alpha \beta }=\partial _{\mu }A_{\nu \alpha \beta }-\partial
_{\nu }A_{\mu \alpha \beta },  \label{33}
\end{equation}%
where

\begin{equation}
A_{\mu \alpha \beta }=\frac{1}{2}(\partial _{\beta }h_{\mu \alpha }-\partial
_{\alpha }h_{\mu \beta }).  \label{34}
\end{equation}%
The expression (33) immediately suggests that $\ R_{\mu \nu \alpha \beta }$
can be seen as an Abelian field strength with $A_{\mu \alpha \beta }=-A_{\mu
\beta \alpha }$ as the gauge potential. In fact, as it is mentioned in Refs.
[1] \ and [26], this interpretation is reinforced by noticing that $R_{\mu
\nu \alpha \beta }$ is invariant under the gauge transformation

\begin{equation}
\delta A_{\mu \alpha \beta }=\partial _{\mu }\lambda _{\alpha \beta },
\label{35}
\end{equation}%
where $\lambda _{\alpha \beta }=-\lambda _{\beta \alpha }$ is an arbitrary
two-form. Now, it is not difficult to prove that, up to surface term, the
action (32) can be written as [1]

\begin{equation}
S_{VII}=\frac{1}{2}\int d^{4}x{}\varepsilon ^{\mu \nu \alpha \beta }{}\Omega
_{\mu \nu }^{\tau \lambda }{}R_{\alpha \beta }^{\sigma \rho }{}\varepsilon
_{\tau \lambda \sigma \rho }.  \label{36}
\end{equation}%
Here, $\Omega _{\mu \nu }^{\alpha \beta }$ is given by%
\begin{equation}
\Omega _{\mu \nu }^{\alpha \beta }=\delta _{\mu }^{\alpha }h_{\nu }^{\beta
}-\delta _{\mu }^{\beta }h_{\nu }^{\alpha }-\delta _{\nu }^{\alpha }h_{\mu
}^{\beta }+\delta _{\nu }^{\beta }h_{\mu }^{\alpha }.  \label{37}
\end{equation}

Suppose we add to the action (36) the topological term

\begin{equation}
S_{T}=\frac{1}{4}\int d^{4}x{}\varepsilon ^{\mu \nu \alpha \beta }{}R_{\mu
\nu }^{\tau \lambda }{}R_{\alpha \beta }^{\sigma \rho }{}\varepsilon _{\tau
\lambda \sigma \rho }  \label{38}
\end{equation}%
and the cosmological constant term

\begin{equation}
S_{C}=\frac{1}{4}\int d^{4}x{}\varepsilon ^{\mu \nu \alpha \beta }{}\Omega
_{\mu \nu }^{\tau \lambda }{}\Omega _{\alpha \beta }^{\sigma \rho
}{}\varepsilon _{\tau \lambda \sigma \rho }.  \label{39}
\end{equation}%
What we obtain is the generalized action [1];

\begin{equation}
S_{VIII}=\frac{1}{4}\int d^{4}x{}\varepsilon ^{\mu \nu \alpha \beta
}{}Q_{\mu \nu }^{\tau \lambda }{}Q_{\alpha \beta }^{\sigma \rho
}{}\varepsilon _{\tau \lambda \sigma \rho },  \label{40}
\end{equation}%
where $Q_{\mu \nu }^{\alpha \beta }$ is defined by

\begin{equation}
Q_{\mu \nu }^{\alpha \beta }=R_{\mu \nu }^{\alpha \beta }+\Omega _{\mu \nu
}^{\alpha \beta }.  \label{41}
\end{equation}%
Moreover, it is not difficult to prove that the action (40) is reduced to
(see Ref. [1] for details)

\begin{equation}
\begin{array}{c}
S_{VIII}=\frac{1}{4}\int d^{4}x{}\varepsilon ^{\mu \nu \alpha \beta
}{}R_{\mu \nu }^{\tau \lambda }{}R_{\alpha \beta }^{\sigma \rho
}{}\varepsilon _{\tau \lambda \sigma \rho }+8\int d^{4}x{}h^{\mu \nu
}(R_{\mu \nu }-\frac{1}{2}\eta _{\mu \nu }R) \\ 
\\ 
-8\int d^{4}x{}(h^{2}-h^{\mu \nu }h_{\mu \nu }).%
\end{array}
\label{42}
\end{equation}%
We recognize in the second and third terms of (42) the Pauli-Fierz action
for linearized gravity with cosmological constant, while the first term is a
total derivative (Euler topological invariant or Gauss-Bonnet term). Note
that the usual cosmological factor $\Lambda $ in the third term can be
derived simply by changing $\Omega \rightarrow a^{2}\Omega ,$ where $a$ is a
constant, and rescaling the total action $S_{VII}\rightarrow \frac{1}{4}%
\Lambda ^{-1}S_{VII},$ with $\Lambda =a^{2}.$

In order to develop a $S$-dual linearized gravitational action we generalize
the action (40) as follows;%
\begin{equation}
S_{IX}=\frac{1}{2}(\lambda ^{+})\int d^{4}x{}\varepsilon ^{\mu \nu \alpha
\beta }{}^{+}Q_{\mu \nu }^{\tau \lambda }{}^{+}Q_{\alpha \beta }^{\sigma
\rho }{}\varepsilon _{\tau \lambda \sigma \rho }+\frac{1}{2}(\lambda
^{-})\int d^{4}x{}\varepsilon ^{\mu \nu \alpha \beta }{}^{-}Q_{\mu \nu
}^{\tau \lambda }{}^{-}Q_{\alpha \beta }^{\sigma \rho }{}\varepsilon _{\tau
\lambda \sigma \rho },  \label{43}
\end{equation}%
where $\lambda ^{+}$ and $\lambda ^{-}$ are two different constant
parameters (playing the analogue role of the parameters $\tau ^{+}$ and $%
\tau ^{-}$ in the Maxwell case) and $^{\pm }Q_{\mu \nu }^{\alpha \beta }$ is
given by%
\begin{equation}
^{\pm }Q_{\mu \nu }^{\alpha \beta }=(\frac{1}{2})^{\pm }N_{\tau \lambda
}^{\alpha \beta }Q_{\mu \nu }^{\tau \lambda },  \label{44}
\end{equation}%
where 
\begin{equation}
^{\pm }N_{\tau \lambda }^{\alpha \beta }=\frac{1}{2}(\delta _{\tau \lambda
}^{\alpha \beta }\mp i\varepsilon _{\,\quad \tau \lambda }^{\alpha \beta }).
\label{45}
\end{equation}%
It turns out that $^{+}Q_{\mu \nu }^{\alpha \beta }$ is self-dual, while $%
^{-}Q_{\mu \nu }^{\alpha \beta }$ is anti self-dual curvature tensors.
Therefore, the action (43) describes self-dual and anti-self-dual linearized
gravity.

Following the steps of section 3 let us introduce a two-form $G$ and use it
for defining%
\begin{equation}
H_{\mu \nu }^{\alpha \beta }\equiv Q_{\mu \nu }^{\alpha \beta }-G_{\mu \nu
}^{\alpha \beta }.  \label{46}
\end{equation}%
We assume that $G_{\mu \nu }^{\alpha \beta }$ satisfies the same indices
symmetry properties as $R_{\mu \nu }^{\alpha \beta }$, namely 
\begin{equation}
\begin{array}{c}
G_{\mu \nu \alpha \beta }=-G_{\mu \nu \beta \alpha }=-G_{\nu \mu \alpha
\beta }=G_{\alpha \beta \mu \nu }, \\ 
\\ 
G_{\mu \nu \alpha \beta }+G_{\mu \beta \nu \alpha }+G_{\mu \alpha \beta \nu
}=0.%
\end{array}
\label{47}
\end{equation}

Now, consider the extended action 
\begin{equation}
\begin{array}{cc}
S_{X}= & \frac{1}{2}(\lambda ^{+})\int dx^{4}\varepsilon ^{\mu \nu \alpha
\beta }{}^{+}H_{\mu \nu }^{\tau \lambda }{}^{+}H_{\alpha \beta }^{\sigma
\rho }{}\varepsilon _{\tau \lambda \sigma \rho }+\frac{1}{2}(\lambda
^{-})\int dx^{4}\varepsilon ^{\mu \nu \alpha \beta }{}^{-}H_{\mu \nu }^{\tau
\lambda }{}^{-}H_{\alpha \beta }^{\sigma \rho }{}\varepsilon _{\tau \lambda
\sigma \rho } \\ 
&  \\ 
& +\int d^{4}x\varepsilon ^{\mu \nu \tau \lambda }{}^{+}W_{\mu \nu }^{\alpha
\beta }{}^{+}G_{\tau \lambda }^{\sigma \rho }{}\varepsilon _{\alpha \beta
\sigma \rho }-\int d^{4}x\varepsilon ^{\mu \nu \tau \lambda }{}^{-}W_{\mu
\nu }^{\alpha \beta }{}^{-}G_{\tau \lambda }^{\sigma \rho }{}\varepsilon
_{\alpha \beta \sigma \rho },%
\end{array}
\label{48}
\end{equation}%
where $W_{\mu \nu \alpha \beta }=\partial _{\mu }V_{\nu \alpha \beta
}-\partial _{\nu }V_{\mu \alpha \beta }$ is the dual field strength
satisfying the Dirac quantization law%
\begin{equation}
\int W\in 2\pi \mathbf{Z.}  \label{49}
\end{equation}%
It is not difficult to see that, beyond the gauge invariance $A\rightarrow
A-d\lambda ,$ $G\rightarrow G$, the partition function

\begin{equation}
Z=\int d^{+}G{}d^{-}G{}dA{}dh{}dV{}e^{-S_{X}}  \label{50}
\end{equation}%
is invariant under%
\begin{equation}
A\rightarrow A+B\;and\;G\rightarrow G+dB,  \label{51}
\end{equation}%
where $B_{\mu \alpha \beta }=-B_{\mu \beta \alpha }$ is an arbitrary tensor.

Starting from (48) one can proceed in two different ways. For the first
possibility, we note that the path integral that involves $V$ is 
\begin{equation}
\int DV\exp (\int d^{4}x\varepsilon ^{\mu \nu \tau \lambda }{}^{+}W_{\mu \nu
}^{\alpha \beta }{}^{+}G_{\tau \lambda }^{\sigma \rho }{}\varepsilon
_{\alpha \beta \sigma \rho }-\int d^{4}x\varepsilon ^{\mu \nu \tau \lambda
}{}^{-}W_{\mu \nu }^{\alpha \beta }{}^{-}G_{\tau \lambda }^{\sigma \rho
}{}\varepsilon _{\alpha \beta \sigma \rho }).  \label{52}
\end{equation}%
Integrating over the dual connection $V$, we get a delta function setting $%
dG=0.$ Thus, using the gauge invariance (51), we may gauge $G$ to zero,
reducing (48) to the original action (43). Therefore, the actions (48) and
(43) are, in fact, classically equivalents.

For the second possibility, we note that the gauge invariance (51) enables
to fix a gauge with $A=0.$ (It is important to note that, at this stage, we
are considering $A_{\mu \alpha \beta }$ and $h_{\mu \nu }$ as independent
fields.) The action (48) is then reduced to 
\begin{equation}
\begin{array}{cc}
S_{X}= & \frac{1}{2}(\lambda ^{+})\int dx^{4}\varepsilon ^{\mu \nu \alpha
\beta }{}^{+}P_{\mu \nu }^{\tau \lambda }{}^{+}P_{\alpha \beta }^{\sigma
\rho }{}\varepsilon _{\tau \lambda \sigma \rho }+\frac{1}{2}(\lambda
^{-})\int dx^{4}\varepsilon ^{\mu \nu \alpha \beta }{}^{-}P_{\mu \nu }^{\tau
\lambda }{}^{-}P_{\alpha \beta }^{\sigma \rho }{}\varepsilon _{\tau \lambda
\sigma \rho } \\ 
&  \\ 
& +\int d^{4}x\varepsilon ^{\mu \nu \tau \lambda }{}^{+}W_{\mu \nu }^{\alpha
\beta }{}^{+}G_{\tau \lambda }^{\sigma \rho }{}\varepsilon _{\alpha \beta
\sigma \rho }-\int d^{4}x\varepsilon ^{\mu \nu \tau \lambda }{}^{-}W_{\mu
\nu }^{\alpha \beta }{}^{-}G_{\tau \lambda }^{\sigma \rho }{}\varepsilon
_{\alpha \beta \sigma \rho },%
\end{array}
\label{53}
\end{equation}%
where 
\begin{equation}
P_{\mu \nu }^{\tau \lambda }\equiv \Omega _{\mu \nu }^{\tau \lambda }-G_{\mu
\nu }^{\tau \lambda }.  \label{54}
\end{equation}

By eliminating $G_{\mu \nu }^{\tau \lambda }$ in (53) we get the dual action

\begin{equation}
S_{XI}=\frac{1}{2}(-\frac{1}{\lambda ^{+}})\int dx^{4}\varepsilon ^{\mu \nu
\alpha \beta }{}^{+}\Xi _{\mu \nu }^{\tau \lambda }{}^{+}\Xi _{\alpha \beta
}^{\sigma \rho }{}\varepsilon _{\tau \lambda \sigma \rho }+\frac{1}{2}(-%
\frac{1}{\lambda ^{-}})\int dx^{4}\varepsilon ^{\mu \nu \alpha \beta
}{}^{-}\Xi _{\mu \nu }^{\tau \lambda }{}^{-}\Xi _{\alpha \beta }^{\sigma
\rho }{}\varepsilon _{\tau \lambda \sigma \rho },  \label{55}
\end{equation}%
Here, $\Xi _{\mu \nu }^{\tau \lambda }$ means

\begin{equation}
\Xi _{\mu \nu }^{\alpha \beta }=W_{\mu \nu }^{\alpha \beta }+\Omega _{\mu
\nu }^{\alpha \beta }.  \label{56}
\end{equation}%
Observe that the complex parameter $\lambda $ has been exchanged by $-\frac{1%
}{\lambda }$ as expected.

\bigskip\ 

\noindent \textbf{6.- A relation between }$F$\textbf{-duality and }$S$%
\textbf{-duality for an Abelian gauge field}

\smallskip\ 

One of our main goals is to establish, in section 7, a possible link between
the $F$-duality and the $S$-duality for linearized gravity. But we shall
first investigate a possible connection between $F$-duality and $S$-duality
in the context of an Abelian gauge field theory.

As we mentioned in section 2, the Maxwell action (7) is not invariant under
the infinitesimal transformations (4) and (5) in spite of the field
equations (2) and (3) are. This problem can be overcome if one solves (3) in
terms of the relation

\begin{equation}
F_{\mu \nu }=\partial _{\mu }A_{\nu }-\partial _{\nu }A_{\mu },  \label{57}
\end{equation}%
and considers canonical variations of the potential $\delta A_{\mu }$
instead of variations of the field strength $\delta F_{\mu \nu }.$ In turn,
in order to maintain duality invariance at the level of the corresponding
canonical action, this forces to introduce what is called superpotential
(see Refs. [5] and [28] for details). However, in this case we are already
using the field equations (3) which, in principle, can not be obtained from
the original action (7). This means that the action (7) needs to be properly
modified in such a way that the field equations (3) are a consequence of an
extended action. The procedure is well known, one introduces an auxiliary
vector field Lagrange multiplier $V^{\mu }$ and writes the new action as

\begin{equation}
S=\frac{1}{2}\int d^{4}xF^{\mu \nu }F_{\mu \nu }+\int d^{4}x\varepsilon
^{\mu \nu \alpha \beta }V_{\mu }\partial _{\nu }F_{\alpha \beta }.
\label{58}
\end{equation}%
Here, of course we are not assuming the form (57) for $F_{\mu \nu }$,
otherwise the second term in (58) is identically zero. In fact, starting
with (58) one can proceed in two different ways. In the first case, varying $%
V_{\mu }$ one obtains the field equation (3) which has the solution (57).
Substituting (57) into the second term of (58) one sees that the action (7)
is recovered. In the second case, it is first convenient to make an
integration by parts obtaining (up to surface term)

\begin{equation}
S=\frac{1}{2}\int d^{4}xF^{\mu \nu }F_{\mu \nu }+\frac{1}{2}\int
d^{4}x\varepsilon ^{\mu \nu \alpha \beta }W_{\mu \nu }F_{\alpha \beta },
\label{59}
\end{equation}%
where $W_{\mu \nu }=\partial _{\mu }V_{\nu }-\partial _{\nu }V_{\mu }$ and
then solving for $F_{\mu \nu }$. In this way, we obtain the relation

\begin{equation}
F^{\mu \nu }=-\frac{1}{2}\varepsilon ^{\mu \nu \alpha \beta }W_{\alpha \beta
}=-^{\ast }W^{\mu \nu },  \label{60}
\end{equation}%
which can be used to get the dual action%
\begin{equation}
S=\frac{1}{2}\int d^{4}xW^{\mu \nu }{}W_{\mu \nu }.  \label{61}
\end{equation}%
Observe that if one assumes (57) then the second term in (59) is identically
zero. An important change in this procedure arises if one assumes a
nontrivial topology. In this case, the solution (57) of (3) no longer is
true. But the correct expression is

\begin{equation}
F_{\mu \nu }\rightarrow H_{\mu \nu }=\partial _{\mu }A_{\nu }-\partial _{\nu
}A_{\mu }-G_{\mu \nu },  \label{62}
\end{equation}%
where the two-form $G$ is a "string" field associated with a nontrivial
topology, so that $dG=0$. This phenomena can be emphasized if instead of
starting with the action (59) one considers the action

\begin{equation}
S=\frac{1}{2}\int d^{4}x\{H^{\mu \nu }H_{\mu \nu }+\varepsilon ^{\mu \nu
\alpha \beta }W_{\mu \nu }H_{\alpha \beta }\},  \label{63}
\end{equation}%
with

\begin{equation}
H_{\mu \nu }=F_{\mu \nu }-G_{\mu \nu }.  \label{64}
\end{equation}

Note that by assuming the relation (62) the action (63) is reduced to

\begin{equation}
S=\frac{1}{2}\int d^{4}x\{H^{\mu \nu }H_{\mu \nu }-\varepsilon ^{\mu \nu
\alpha \beta }W_{\mu \nu }G_{\alpha \beta }\}.  \label{65}
\end{equation}%
This development leads to the conclusion that rather than looking for the
invariance of the action (7) under the infinitesimal transformation (4) one
should consider invariance of the action (63) or (65) under such
transformations. But one may recognize that the action (65) has exactly the
same form as the expression (19) (see section 3) which was considered in the
context of $S$-duality approach. The main difference between (65) and (19)
is that in (19) one considers $^{\pm }H^{\mu \nu }$, $^{\pm }W_{\mu \nu }$
and $^{\pm }G_{\alpha \beta }$ rather than $H^{\mu \nu }$, $W_{\mu \nu }$
and $G_{\alpha \beta }$ as in (65). Further the parameters $^{\pm }\tau $
are considered in (19), while in (65) this is not the case. This means that
(65) can be considered as a particular case of (19). And in this context one
should expect that invariance of (19) leads to a reduced invariance of (65).
Indeed, the transformation (20), namely $\delta A=B,$ $\delta G=dB,$ where $%
B $ is any one-form, also leaves the action (65) invariant. It is
interesting to note that the infinitesimal transformation (4) can be
considered as a particular case of (20) as soon as one also assumes the
transformation $\delta G=\beta ^{\ast }F$ for the "string" field $G$. One of
our conclusions is that in order to implement the transformation (4) at the
level of the action of the Mawxell theory one needs to introduce an
auxiliary field $G$ and considers (63) or (65) as starting point rather than
(7).

Let us use the notation $D=dB$. From (64) we then observe that

\begin{equation}
\delta H_{\mu \nu }=\delta F_{\mu \nu }-\delta G_{\mu \nu }=D_{\mu \nu
}-D_{\mu \nu },  \label{66}
\end{equation}%
which is of course identically equal to zero. But writing $\delta H_{\mu \nu
}$ as in (66) it suggests to consider (4) $\delta F_{\mu \nu }=\beta ^{\ast
}F_{\mu \nu }$ as a particular case with $D_{\mu \nu }=\beta ^{\ast }F_{\mu
\nu }$ and $\delta G_{\mu \nu }=\beta ^{\ast }F_{\mu \nu }$. In fact, this
possibility seems to pass unnoticed before in the context of $S$-duality
formalism. Perhaps because the invariance of (66) was written in terms of $%
\delta A_{\mu }$ rather than in terms of $\delta F_{\mu \nu }$. It is true
that $\delta A_{\mu }$ implies $\delta F_{\mu \nu }$ but the converse is no
in general true; unless one considers nonlocal formalism in the sense $%
\delta A=d^{-1}D$, which in the case of the variation $\delta F_{\mu \nu
}=\beta ^{\ast }F_{\mu \nu }$ means $\delta A=B=\beta d^{-1\ast }F$. It is
tempted to assume that from the canonical point of view this is equivalent
to introduce what is called superpotential [5, 31]. In other words, our
conjecture is that the "string" field $G$ and the superpotential are closely
related [35].

\bigskip\ 

\noindent \textbf{7.- }$F$\textbf{-duality and }$S$\textbf{-duality in
linearized gravity}

\smallskip\ 

An application of the prescription of the previous section to the case of
linearized gravity is straightforward. From (46) one sees that $H_{\mu \nu
}^{\alpha \beta }=Q_{\mu \nu }^{\alpha \beta }-G_{\mu \nu }^{\alpha \beta }$
remains invariant under the transformations

\begin{equation}
\begin{array}{c}
\delta Q_{\mu \nu }^{\alpha \beta }=D_{\mu \nu }^{\alpha \beta }, \\ 
\\ 
\delta G_{\mu \nu }^{\alpha \beta }=D_{\mu \nu }^{\alpha \beta }.%
\end{array}
\label{67}
\end{equation}%
Here, $D_{\mu \nu }^{\alpha \beta }$ is and arbitrary two-form with the
property $D=dB$, where $B$ is any "one-form". This implies that the action
(48) is invariant under (67).

As a particular case of (67) one writes%
\begin{equation}
\delta Q_{\mu \nu }^{\alpha \beta }=\beta ^{\ast }Q_{\mu \nu }^{\alpha \beta
}.  \label{68}
\end{equation}%
This corresponds to consider $D_{\mu \nu }^{\alpha \beta }=\beta ^{\ast
}Q_{\mu \nu }^{\alpha \beta }$. The expression (68) refers of course to
infinitesimal rotations and therefore we have found a mechanism to make the
extended action (48) invariant under such rotations. Again, one can try to
relate (68) with the gauge field $A_{\nu \alpha \beta }$ according to (33)
but this would imply a nonlocal variation $\delta A=\beta d^{-1\ast }Q_{\mu
\nu }^{\alpha \beta }$. It is intriguing that with this procedure we do not
even need to consider the perturbation $h_{\mu \alpha }$ as in the canonical
method of Ref. [5]. However, one should expect that if the action (48) is
written in a canonical form a link between what it is called a
superpotential in Ref. [5] and the auxiliary field $G_{\mu \nu }^{\alpha
\beta }$ must be found.

\bigskip\ 

\bigskip\ 

\noindent \textbf{8.- Discussion and final comments}

\smallskip\ 

In this work we have shown that the $F$-duality is indeed contained in the $%
S $-duality formalism as proposed in the Ref. [30]. One of the advantage of
this identification is that it is not necessary to rely in the canonical
formalism in order to implement duality invariance at the level of the
action. In a sense $S$-duality provides the route that it is necessary to
follow in the case of the $F$-duality program. In fact, $S$-duality
establishes that duality can be achieved at the level of the action by
adding a $\theta $ term to the Maxwell action and by introducing an
auxiliary two form $G$. It turns out that this is also true for linearized
gravity as we have pointed out in section 7.

These results also suggests to consider the coupling parameter $\tau $ in
the $F$-duality formalism. This is because the partition function $Z(\tau )$
in the $S$-duality approach has the property $Z(\tau )=Z(-\frac{1}{\tau })$
or $Z(\lambda )=Z(-\frac{1}{\lambda })$ as it can be deduced from our
discussion of section 3 and 5, respectively. In fact, writing symbolically

\begin{equation}
Z(\tau )=\int \exp (iS_{IV}),  \label{69}
\end{equation}%
where $S_{IV}$ is given in (19), for the case of Maxwell theory and

\begin{equation}
Z(\lambda )=\int \exp (iS_{X}),  \label{70}
\end{equation}%
where $S_{X}$ is given in (48), for the case of linearized gravity, from the
results of section 3 we may establish that (69) has the two limits

\begin{equation}
\int \exp (iS_{III})\leftarrow \int \exp (iS_{IV})\rightarrow \int \exp
(iS_{V}),  \label{71}
\end{equation}%
(where $S_{III}$ and $S_{V}$ are given by (12) and (22), respectively),
while from the discussion of section 5 we may establish that (70) gives

\begin{equation}
\int \exp (iS_{IX})\leftarrow \int \exp (iS_{X})\rightarrow \int \exp
(iS_{XI}),  \label{72}
\end{equation}%
(where $S_{IX}$ and $S_{XI}$ are given by (43) and (55), respectively).
Therefore, one finds that (71) and (72) imply the symmetries $Z(\tau )=Z(-%
\frac{1}{\tau })$ and $Z(\lambda )=Z(-\frac{1}{\lambda })$ respectively.

It has been shown [30] that $Z(\tau )$ also contains the symmetry $Z(\tau
)=Z(\tau +1)$ showing with this that $Z(\tau )$ is symmetric under the full
group $SL(2,Z)$. So, it may appear interesting to see whether $F$-duality
formalism may also be connected with the transformation $\tau \rightarrow
\tau +1$. In what follow we shall outline this possibility.

First we note that if we consider the infinitesimal transformations (4) and
(5) we find that the self-dual (antiself-dual) field strength transforms as

\begin{equation}
\delta ^{\pm }F^{\alpha \beta }=\pm i\beta ^{\pm }F^{\alpha \beta }.
\label{73}
\end{equation}%
Therefore, we discover that the action (12) transforms as

\begin{equation}
\delta S_{III}=i\beta \left\{ (\tau ^{+})\int d^{4}x{}^{+}F^{\mu \nu
}{}^{+}F_{\mu \nu }-(\tau ^{-})\int d^{4}x{}^{-}F^{\mu \nu }{}^{-}F_{\mu \nu
}\right\} .  \label{74}
\end{equation}%
In this case we have left the parameters $\tau ^{+}$ and $\tau ^{-}$
unchanged. However, we can obtain similar result if we leave the field
strength $F^{\alpha \beta }$ unchanged and we require the parameters $\tau
^{+}$ and $\tau ^{-}$ transform as follows

\begin{equation}
\begin{array}{c}
\tau ^{\prime +}=\tau ^{+}+i\beta \tau ^{+}, \\ 
\\ 
\tau ^{\prime -}=\tau ^{-}-i\beta \tau ^{-}.%
\end{array}
\label{75}
\end{equation}%
An interesting possibility arises if one considers the particular cases $%
\beta =\frac{1}{\tau ^{+}}$ or $\beta =\frac{1}{\tau ^{-}},$ leading in any
case to the result

\begin{equation}
\begin{array}{c}
\tau ^{\prime +}=\tau ^{+}+i, \\ 
\\ 
\tau ^{\prime -}=\tau ^{-}-i,%
\end{array}
\label{76}
\end{equation}%
which is similar to the expected form $\tau \rightarrow \tau +1.$

The result (74) means that the action (12) is no invariant under (73) or
(75). However, if one considers the transformations (76) this is not
necessarily true for the associated partition function $Z=Z(\tau ^{\pm })$,
namely $Z(\tau ^{\pm })=\int \exp (iS_{III})$. In fact the reason for this
is that using (76) one discovers that the expression (74) becomes

\begin{equation}
\delta S_{III}=i\left\{ \int d^{4}x{}^{+}F^{\mu \nu }{}^{+}F_{\mu \nu }-\int
d^{4}x{}^{-}F^{\mu \nu }{}^{-}F_{\mu \nu }\right\} ,  \label{77}
\end{equation}%
which can be reduced to the $\theta $ term

\begin{equation}
\delta S_{III}=\theta \int d^{4}x{}F^{\mu \nu }{}^{\ast }F_{\mu \nu }.
\label{78}
\end{equation}%
Since from (13) we have $\tau =\frac{1}{g^{2}}+i\theta $ one obtains $\delta
\tau =i\delta \theta $ and therefore the prescription (76) implies $\delta
\theta =1$ which means

\begin{equation}
\theta \rightarrow \theta +1.  \label{79}
\end{equation}%
So, by assuming the smallest possible value for $\int d^{4}x{}F^{\mu \nu
}{}^{\ast }F_{\mu \nu }$ one may recognize that the term $exp(\delta
S_{III}) $ leaves the partition function $Z=Z(\tau ^{\pm })$ invariant.

In references [36]-[38] it is also discussed a kind of $F$-duality from the
point of view of field equations rather than actions. For new directions of
research it may be interesting to establish the precise relations of such a
references with our formalism.

Finally, in references [30] and [39] it is explained that the action (12) is
invariant mod \ $2\pi n$ no only under the change $\tau \rightarrow \tau +1$
when $M$ is an spin manifold but also under the change $\tau \rightarrow
\tau +2$ for any a closed four manifold $M$. It may be interesting for
further research to explore what this means in both scenarios; Maxwell
theory and linearized gravity.

\bigskip\ 

\noindent \textbf{Acknowledgments: }J. A. Nieto would like to thank to L.
Ruiz and J. Silvas for helpful comments. This work was partially supported
by grants PIFI 3.2 and 3.3.

\smallskip\

\end{document}